# COHERENT SYNCHROTRON RADIATION MEASUREMENTS IN THE CLIC TEST FACILITY (CTF II)

H.H. Braun, R. Corsini, L. Groening, F. Zhou, CERN, Geneva, Switzerland
A. Kabel, T. Raubenheimer, SLAC, Menlo Park, CA 94025, USA
R. Li, TJNAF, Newport News, VA 23606, USA
T. Limberg, DESY, Hamburg, Germany

*Abstract*

Bunches of high charge (up to 10 nC) are compressed in length in the CTF II magnetic chicane to less than 0.2 mm rms. The short bunches radiate coherently in the chicane magnetic field, and the horizontal and longitudinal phase space density distributions are affected. This paper reports the results of beam emittance and momentum measurements. Horizontal and vertical emittances and momentum spectra were measured for different bunch compression factors and bunch charges. In particular, for 10 nC bunches, the mean beam momentum decreased by about 5% while the FWHM momentum spread increased from 5% to 19%. The experimental results are compared with simulations made with the code TraFiC[4].

## 1 INTRODUCTION

Short electron bunches traversing a dipole with bending radius $\rho$ can emit coherent synchrotron radiation (CSR) at wavelengths longer than the bunch length. The enhancement of the radiated power, with respect to the classical synchrotron radiation, can be expressed as [1]:

$$\Delta P_{coh} \approx 0.028\, N^2\, \frac{c\, e^2}{\varepsilon_0\, \rho^{2/3}\, \sigma_Z^{4/3}}$$

where a longitudinal Gaussian distribution of the $N$ particles with constant rms bunch length $\sigma_Z$ along the curved trajectory are assumed.

The CSR induces an average momentum loss and a momentum spread on the bunch. For relativistic beams the momentum loss is independent of the beam energy and the effect can be treated like a wake-field. Since it takes place in a dispersive region, the transverse phase space distribution is also affected, and the beam emittance in the bending plane increases.

These effects are a concern in all accelerator applications in which high-charge short bunches are needed, e.g., free electron lasers, linear colliders and two-beam accelerators. Analytical treatments have been developed for idealised conditions [1-4] but for many practical cases the effect must be treated numerically. Some codes have been developed for this purpose [5,6], but a complete benchmark of the codes with measurements is so far lacking. The parameters of the drive beam in the second Compact Linear Collider Test Facility, CTF II [7] are well suited to study the CSR effect experimentally.

CTF II (see Fig. 1) was built to demonstrate the feasibility of two-beam acceleration at 30 GHz. The high-charge drive beam is generated in a 3 GHz RF gun, and accelerated in two Travelling Wave Structures (TWS). The drive beam bunches are then compressed in a magnetic chicane, and generate 30 GHz RF power in a series of Power Extraction and Transfer Structures (PETS). This power is used to accelerate a low-charge probe beam, in a parallel beam line.

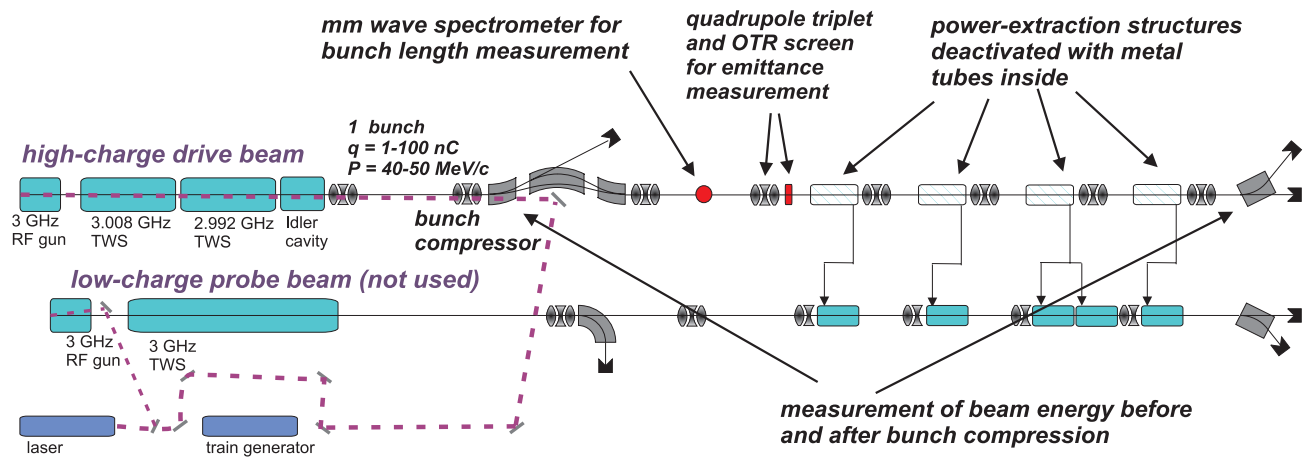

Figure 1: The CLIC Test Facility CTF II. The CSR experiments were performed in the drive beam line (upper part).

Only the drive beam line of CTF II has been used to perform the CSR experiment. Single bunches with charges of 5 and 10 nC have been used. For both charges, the horizontal and vertical beam emittances and the momentum spectra of the drive beam have been measured as a function of the deflection angle in the chicane. Similar measurements have been performed already in CTF II [8], but limited to the emittances. The results of simulations made using the code TraFiC$^4$ [5] are reported for comparison.

## 2 EXPERIMENTAL SETUP

Bunch compression is used in CTF II to enhance the 30 GHz power production, and is achieved by acceleration off-crest (such that particles at the tail of the bunch have higher energies than those at the head), in combination with a magnetic chicane composed of three rectangular dipoles. The deflection angle in the first and the last dipole can be varied from 3.7° to 14° (twice these values in the central dipole). The corresponding energy dependence of the path length $R_{56} = ds / (dE/E)$ ranges from 6 mm to 90 mm. Due to the construction of the vacuum chamber, the chicane cannot be switched off completely. Using this scheme, rms bunch lengths of less than 0.4 mm can by achieved for bunch charges of 10 nC.

The beam momentum spectra were measured at the entrance to the chicane by switching off the last two chicane magnets and using the first one as a spectrometer. A second spectrometer at the end of the line permits the measurement of the beam spectra after the passage in the chicane. The PETS were deactivated by shielding them with metallic tubes, in order not to perturb the momentum distribution. Two wall-current monitors placed in front of the chicane and in front of the second spectrometer measured the beam intensities, in order to monitor the losses.

The bunch lengths were measured behind the chicane by analysing the frequency spectrum of the mm-wave radiation excited by the beam passing an RF waveguide connected to the vacuum chamber [9]. Transverse beam profiles were recorded after the chicane using an optical transition radiation (OTR) screen and a camera. The horizontal and vertical emittances were measured simultaneously by using the quadrupole scanning technique. In order to cover the full range of phase advances from 0° to 180° in both transverse planes, the quadrupole strengths of a bipolar triplet were varied independently. For accurate determination of the beam widths, only those profiles were selected which fitted on the OTR screen.

The off-crest RF phase and the range of chicane settings were selected in order to achieve over-compression of the initial bunch length, thus covering a sufficient range in bunch lengths.

## 3 MOMENTUM SPECTRA AND BUNCH LENGTH

The beam momentum spectra measured for different bunch compressor settings are shown in Fig. 2 (5 nC case) and Fig. 3 (10 nC case). The spectra from TraFiC$^4$ simulations are also shown. Since the spectrometer at the end of the beam line consists of a deflecting dipole it also represents a source of CSR, and the spectra result from the sum of the CSR effects in the compressor chicane and in the spectrometer.

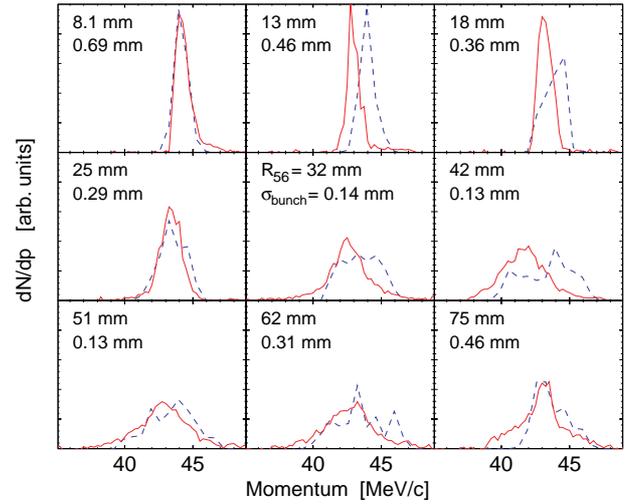

Figure 2: Measured (solid line) and calculated (dotted line) momentum spectra at the end of the beam line for different compressor settings $R_{56}$, with a bunch charge of 5 nC. The measured bunch lengths are also shown (lower values).

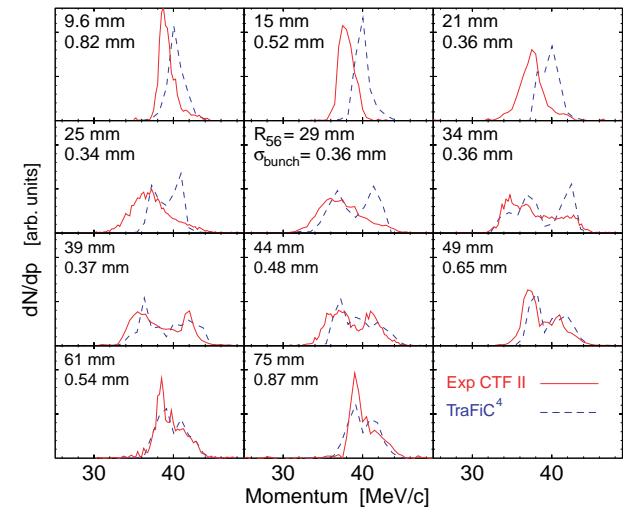

Figure 3: Measured (solid line) and calculated (dotted line) momentum spectra at the end of the beam line for different compressor settings $R_{56}$, with a bunch charge of 10 nC.

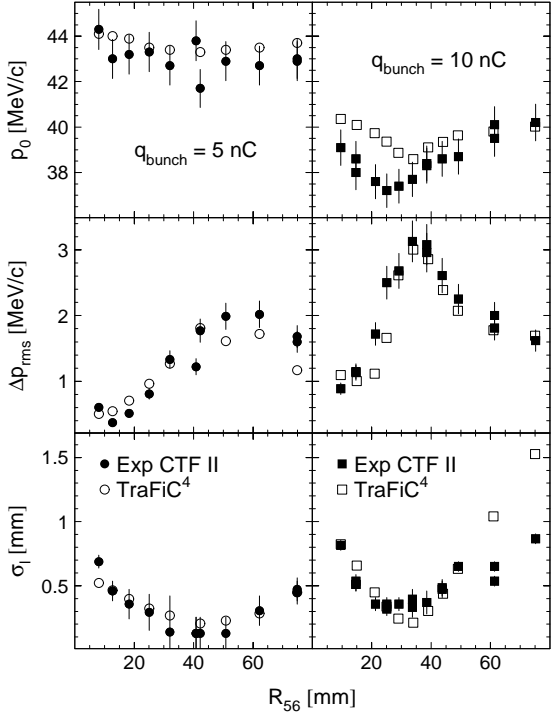

Figure 4: Measured and calculated rms bunch lengths, mean momenta and momentum spreads as functions of $R_{56}$, for bunch charges of 5 nC (left) and 10 nC (right).

As the bunches are compressed, the spectrum broadens, and in the 10 nC case it develops two maxima beyond maximum compression. During over-compression the spectrum narrows again.

The measured bunch lengths and the mean momenta and rms momentum spreads calculated from the measured spectra are plotted in Fig. 4, as function of $R_{56}$. They are compared with the results of the TraFiC$^4$ simulations.

For 5 nC, no significant changes of the mean beam momentum was observed, although the momentum spread increased by a factor four at full compression with respect to the initial spread, and decreased at over-compression. For charges of 10 nC, a decrease of the mean momentum by 2 MeV/c, i.e. 5% of the initial beam momentum, was measured at full compression. The momentum spread shows a behaviour similar to the one described for 5 nC. In the 5 nC case the dependence of the measured bunch length on $R_{56}$ shows a symmetric shape as expected from linear longitudinal dynamics. On the other hand, the asymmetric shape of the corresponding curve for 10 nC indicates a strong impact of CSR on the longitudinal properties for $R_{56} > 20$ mm.

Using the program TraFiC$^4$, the bunch, consisting of 500 macro particles, was tracked from the entrance to the chicane to the end of line spectrometer dipole. At each step, the sum of fields resulting from magnetic fields, space charge and CSR wake-fields was applied to the macro particles. The resulting bunch lengths, emittances and momentum spectra have been calculated from the final distributions.

In order to determine the initial conditions to be used in the simulations, the bunch lengths measured after the chicane have been used together with the momentum spectra recorded at the entrance to the chicane. The initial phase space distributions in the longitudinal plane have been reconstructed taking into account the contribution to the intra-bunch momentum correlation of the off-crest RF phase and of the short-range longitudinal wake-fields in the accelerating structures. The initial bunch length has been chosen to fit the momentum spectra at the chicane entrance and the measured bunch lengths, assuming that the CSR does not influence the bunch length. As already mentioned above, this assumption seems to be justified only for $R_{56} < 20$ mm in the 10 nC case. Indeed, while a linear longitudinal model describes well the measured bunch length dependence from $R_{56}$ for 5 nC, no set of initial conditions could be found that can describe the measured values for 10 nC. Therefore only the bunch lengths for small $R_{56}$ have been used in this case. Since only a few measurement points are available for the fit, the input parameters for the 10 nC measurements are known with less accuracy.

In the case of 5 nC the dependence of the observables on $R_{56}$ was well reproduced. For 10 nC the calculated bunch lengths agreed with the measured ones until full compression, but the asymmetric shape of the bunch length curve was not reproduced, although the TraFiC$^4$ simulation looks more accurate than the simple linear calculation. The amount of momentum loss for 10 nC was calculated correctly, but the experimental and theoretical curves are shifted with respect to each other, possibly indicating different initial longitudinal beam parameters in the experiments and the simulations. Another possible explanation is a fluctuation in the initial beam parameters while the measurement was taken. Especially, variations in RF power and phase could be invoked, since the same initial conditions have been used to simulate all of the measurement points and a phase variation of a few degrees would be enough to explain the discrepancies found.

On the other hand, a very good matching of the measured and calculated momentum spreads was found for both cases. In the case of 5 nC, the simulations reveal no splitting of the maximum and the shapes of the spectra show a good agreement. In the 10 nC case, although the rms widths are in good agreement, the measured and simulated momentum spectra have somewhat different shapes. The formation of two maxima occurs in the simulations for smaller values of $R_{56}$. Since the CSR wake depends on the exact longitudinal charge distribution within the bunch, the details of the momentum spectrum shape should depend on that as well.

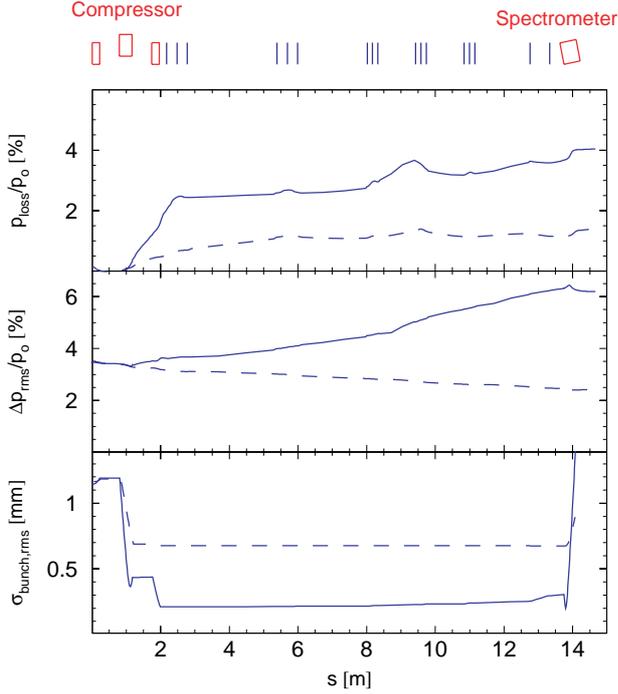

Figure 5: Evolution of the bunch length, the rms momentum spread and the momentum loss (from bottom to top) along the CTF II beam line as simulated with TraFiC$^4$, for 10 nC. The results with two different setting of the chicane are shown, corresponding to $R_{56} = 34$ mm (solid line) and $R_{56} = 15$ mm (dashed line).

In the simulation an initial Gaussian charge distribution has been assumed, but previous bunch length measurements with a streak camera indicate a different, asymmetric charge distribution. The simulations show that the splitting of the maxima in the spectra for 10 nC is due to CSR in the chicane, since such a feature is present in the simulated spectrum right after the chicane.

On the other hand, the broadening of the spectra seems to be caused by space charge during the drift to the spectrometer. This is apparent in Fig. 5, where the evolution of the bunch length, the momentum spread and the momentum loss along the beam line, simulated with TraFiC$^4$, are shown.

In spite of the fact that some features of the simulated momentum loss curve are not yet understood, from Fig. 5, it can be seen that the momentum loss is mainly concentrated in the chicane region where the bunches are short, and is a direct indication of CSR emission. On the other hand, for $R_{56} = 34$ mm, the momentum spread increases along the whole beam line after the chicane, showing that such growth is not directly caused by a CSR effect.

## 4 EMITTANCES

The measured horizontal and vertical emittances after the chicane are plotted in Fig. 6 as function of $R_{56}$, together with the bunch lengths.

The measured vertical emittances for 5 nC are constant within the error bars, and are consistent with the simulation results. The measured horizontal emittances for 5 nC increase until full compression, then a saturation occurs. The four highest values seem to be shifted with respect to the lower values. This shift might be due to dispersion at the location of the OTR screen, caused by a mismatch of the relative magnetic strength of the chicane dipoles. It must be noted that these four points were recorded after switching off the last two magnets of the chicane in order to check the beam momentum, and then switching on the magnets again.

For 10 nC the measured vertical emittance shows a maximum at full compression. This qualitative behaviour is found also in the simulations. The measured data points for the horizontal emittance scatter after over-compression, thus not allowing meaningful conclusions. The huge growth of the horizontal emittance predicted by the simulations was not found experimentally. It can be noted that the increase of the momentum spread also affects the accuracy of the emittance measurements and contributes to the error bars and the scattering of the data points.

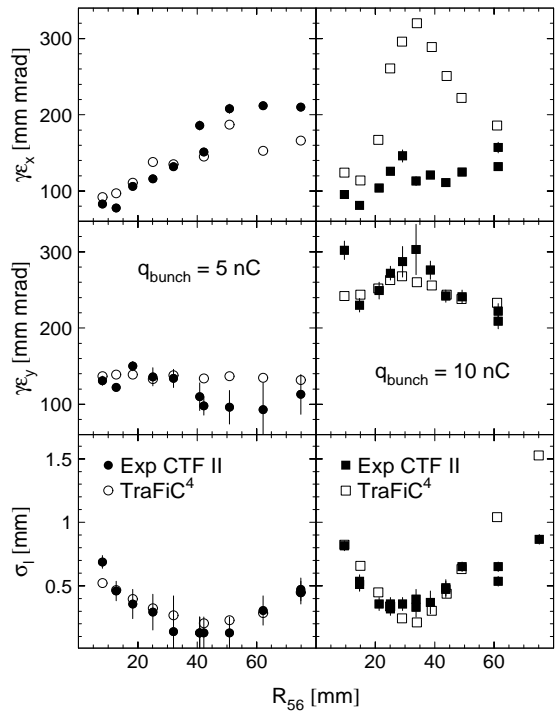

Figure 6: Measured and calculated rms bunch lengths and transverse beam emittances as function of $R_{56}$ for bunch charges of 5 nC (left) and 10 nC (right).

# 5 CONCLUSIONS AND OUTLOOK

The measurements made in the CTF II drive beam line showed clear signs of CSR emission. In particular the momentum loss and bunch length as a function of the bunch compressor setting in the 10 nC case are well explained by the CSR effect, such as the two-peak momentum spectrum shape at 10 nC. A good agreement was found between measurements and simulations made with the TraFiC$^4$ code for all observables, except the emittance growth for bunches of 10 nC, which is not yet understood.

Future experiments will aim in particular at an experimental investigation of the shielding effect of the beam pipe on the CSR emission. The extension of the CSR spectrum is limited at the high-frequency end by the bunch length, and at the low-frequency end by the cut-off of the beam pipe. When the beam pipe dimensions are small with respect to the bunch length, CSR emission is suppressed. While this is indeed the case in many accelerators, in the CTFII chicane the free space approximation is valid. It would be interesting to explore the intermediate regime since often only a partial shielding can be used to reduce the unwanted CSR effect in the case of very small bunches. Several approaches to the calculation of the shielding effect exist [1-4, 10], using different approximations. The TraFiC$^4$ code is also capable of treating the shielding, and could be benchmarked against the measurements.

Short bunches will be passed through a new four-magnet chicane installed downstream of the bunch compressor. Three vacuum chambers of different height will be used in order to provide different shielding environments. The new chicane has been designed for large deflection angles and small dispersion functions, i.e. small $R_{56}$. Therefore the bunch length can be made short and relatively constant in the new chicane, while CSR emission in this region will be high. A further advantage will be the smaller disturbance from possible dispersion errors.